\documentstyle[twoside,fleqn,espcrc2]{article}

\newcommand{\AmS}{{\protect\the\textfont2
  A\kern-.1667em\lower.5ex\hbox{M}\kern-.125emS}}

\title{Flavor, Compositeness, and Dynamical Breaking of Supersymmetry}

\author{A. E. Nelson\address{Department of Physics, Box 351560,
        University of Washington, Seattle, WA 98195-1560, USA}
        \thanks{this research supported in part by the 
DOE under grant \#DE-FG03-96ER40956.}}
       
\begin{document}
\begin{abstract}  We should be taking advantage of recent gains in our
nonperturbative understanding of supersymmetric gauge theories to
find the ``standard'' model of  of dynamical supersymmetry breaking,
and possibly of flavor as well.
As an illustration of the possibilities for understanding the flavor 
hierarchy,  I 
describe a realistic, renormalizable, supersymmetric model 
with a compositeness
scale   of $\sim 1-3$ TeV for the top quark, the left handed bottom
quark, and the
up-type Higgs. The top-Higgs Yukawa coupling is a dynamically
generated strong interaction effect, and is naturally large, while
the other Yukawa couplings are  suppressed. 
\end{abstract}
\maketitle
\section{Introduction: Can SUSY Gauge Dynamics Solve Our Problems?}
 Holomorphy and Duality have taught us a lot of nonperturbative information
about low energy dynamics of N~$=1$
 supersymmetric gauge theories \cite{seiberg}. One might hope that
this would turn out to be useful in understanding some long standing
puzzles in particle physics.
\section{The Gauge Hierarchy Problem}
Dynamical Breaking of Supersymmetry  (DSB) at a scale which is
exponentially small when compared with the Planck scale $m_P$, is a
potentially beautiful solution to the problem of why the weak scale is
so much lower than the Planck scale \cite{witten}. 
Until recently only 4 examples of
DSB were known \cite{ads,akmr}, none of which yielded a realistic
candidate model of particle physics \cite{ads}. We now know of several
new mechanisms and many new  classes of DSB  models 
\cite{iss,dnns,hidden,iy,it,pst,crs,chou,crsl}.
In the last two years 
we have learned that supersymmetry can break dynamically in models with 
classically flat
directions, with non-chiral representations of the gauge group, with
gauge singlet superfields, 
without dynamically generated superpotentials, and without
any $U(1)$~R-symmetry \cite{rsymm}. 

While many of the new DSB models can be
supplemented with additional sectors to yield
realistic theories, no really compelling 
``standard model'' of supersymmetry
breaking has emerged. All the models require the addition of a
MSSM (Minimal Supersymmetric Standard Model) sector to be
realistic. 
Hidden sector models are not renormalizable or
predictive, and do not explain the absence of flavor changing neutral
currents or electric dipole moments, while visible sector gauge
mediated models require that, in addition to the MSSM sector, 
a ``messenger sector'' of
new, heavy,  vector-like quarks and leptons be
tacked on. Still, further exploration might reveal a plausible 
DSB model whose low
energy limit contains the standard model, or at least one with room in
its global symmetry group to embed the standard model gauge
interactions, so that the the messenger sector could be avoided. 

Note that any interpretation of the Fermilab
$ee\gamma\gamma$ event involving decay into a gravitino  
\cite{dns,ddrt,akkmm,dtw,bkw,dine,ddrtw} implies a rather low
($<{\cal O}(100~{\rm TeV}) $ ) supersymmetry breaking scale. Since the 
messenger quarks and leptons should also have mass in the 30--100~TeV
range \cite{dnns}, if this event is a  signal for a 
light gravitino then we
have an indication that the DSB and messenger sectors are the
same. In fact there are several ways to merge the DSB and messenger 
sectors
\cite{progress}.
\section{Flavor}
 Even more puzzling than the supersymmetry breaking mechanism is the
explanation for the  hierarchy  of quark and lepton masses and mixing
angles.
It is intriguing to speculate that strongly coupled
dynamics could lie behind the generational structure.  For instance,
in the context of supersymmetry, at 
least some of the superpotential
couplings of the MSSM might have a dynamical origin. 

A proposal along these lines was made in  \cite{topYukawa} and in 
\cite{highscale} in which a dynamical mechanism for
generating the top quark Yukawa coupling was suggested.  
In this ``quindecuplet''
scenario, a confining $SU(2)_C$ gauge theory,  has as its low energy
limit  a 15 dimensional multiplet of composite particles, containing 
the top quark, left-handed bottom quark, the  up-type Higgs, and the
left handed tau anti-lepton.
 The ordinary $SU(3)_c\times SU(2)_w\times
U(1)_Y$ gauge interactions can be embedded into an  $SU(5)$ global
symmetry of the strong interactions, under which
the composite particles transform as $ {\bf 5}+ {\bf 10}$.  The top
quark Yukawa coupling is generated by a strong coupling effect of
confinement \cite{sutwo} and the bottom quark mass is generated through an
 higher-dimension operator arising from Planck scale physics. 
Viable three-generation models,
employing all or part of this mechanism with the compositeness scale
near to the Planck scale, were proposed in \cite{highscale}.  However, the
compositeness scale must be very  high or proton decay would be  too rapid.
Hence, other than the postdiction of qualitative features of the fermion mass
hierarchy, these models make no predictions. 

Here I would like to describe a version of the quindecuplet theory in which the
proton is stable, and   the compositeness scale of the top quark can be low,
$\sim 1$~TeV. The model is realistic and is a good laboratory to study possible
low energy signals which could arise from compositeness \cite{composite}. The
main difference with the model of \cite{highscale} is that all components of the
$\tau$ lepton  are fundamental particles---there is a composite particle with
the gauge quantum numbers of the left handed tau anti lepton but it must carry
baryon number +1 and so is identified as a new, exotic ``triquark'' particle,
the $\bar E$. One family of quarks and leptons results from the particle content
shown in table~1. The ${\tt\bf N},
{\tt\bf N'},{\tt\bf \bar N},
{\tt\bf \bar N'}$ particles are given large masses---and integrating them out
will result in the nonrenormalible operators responsible for the bottom quark  
mass. The fifteen light composite fields of this model are
\begin{equation}\begin{array}{l}
 q\sim\pmatrix{t\cr b}\sim {\tt\bf d}{\tt\bf h}\quad\quad\quad\bar t\sim{\tt\bf
d}{\tt\bf d}
\quad\quad\quad H\sim {\tt\bf h}{\tt\bf n}\\   D\sim{\tt\bf d}{\tt\bf
n}\quad\quad\quad \bar E\sim {\tt\bf h}{\tt\bf h}\ , \end{array}
\end{equation}
which we identify with the top and left handed bottom quarks, the up-type
Higgs, and  an exotic ``diquark'' and ``triquark''.   To get three~families the
model is triplicated---three different
$SU(2)_C$'s with different compositeness scales are introduced. The quark mass
hierarchy is a result of the 3 different compositeness scales.

\begin{table}
\setlength{\tabcolsep}{0.2pc}
\newlength{\digitwidth} \settowidth{\digitwidth}{\rm 0}
\catcode`?=\active \def?{\kern\digitwidth}
\caption{One Family Composite Model}
\begin{tabular}{crrrr}
\hline
 Preon Field   & $SU(2)_C$&$SU(3)_c$&$SU(2)_w$&$U(1)_y$\\
\hline
${\tt\bf d}$            & $2$ & $3$ & $ 1\ $ & $ -1/3$ \\
${\tt\bf h}$            & $2$ & $1$ & $ 2\ $ & $ 1/2 $ \\
${\tt\bf n},{\tt\bf N},
{\tt\bf N'},{\tt\bf \bar N},
{\tt\bf \bar N'}$        & $2$ & $1$ & $1\ $ & $ 0   $ \\
$\bar d, \bar D$        & $1$ & $3$ & $1\ $ & $1/3  $ \\
$\bar H$, $\ell$        & $1$ & $1$ & $2\ $ & $-1/2$  \\
$E$                     & $1$ & $1$ & $1\ $ & $-1$    \\
$\bar e$                & $1$ & $1$ & $1\ $ & $1$     \\
\hline
\end{tabular}
\end{table}

Here I will briefly summarize how the model reproduces the particle masses and mixing
angles.
\subsection{Higgs, $D$ and $E$ masses}
The tree level superpotential contains the terms
\begin{equation} 
W_{\rm tree}\supset \eta^{H} {\tt\bf h} {\tt\bf n} {\bar H} 
+\eta^{D} {\tt\bf d} {\tt\bf n} \bar D
+\eta^E{\tt\bf h}{\tt\bf h}  E  \ .
\end{equation} The first term gives the infamous ``$\mu$'' Higgs mass term of the
MSSM, of size  $\eta^H$ times the compositeness scale.
Hence unless we assume $\eta^H$ is extremely small the compositeness
scale
$\Lambda$ 
should not be too far above the weak scale. Similarly, the second two terms
result in $D$ and $E$ masses proportional to $\eta^D\Lambda$ and 
$\eta^E\Lambda$.
\subsection{The Top Mass} Below the confinement scale, a superpotential is
generated dynamically for the composite particles \cite{sutwo}
\begin{equation}
W_{\rm dynamical}\propto q\bar t H+ qqD + \bar t D \bar E\ .\end{equation}
The first term, the top-Higgs Yukawa coupling, is a nonperturbative
effect which we expect to be large---implying that $\tan\beta$ could be small. 
Note that the
$D$ must be assigned baryon number
$-2/3$ and the
$E$ carries baryon number of $1$. 
\subsection{The Bottom Mass}
This mass must come from the term
\begin{equation}
W_{\rm effective}\supset {1\over M}{\tt\bf d}{\tt\bf h}\bar b \bar
H\ ,\end{equation} which results in a b-quark Yukawa coupling of order
$\Lambda/M$, where $\Lambda$ is the compositeness scale. This nonrenormalizable
term results from integrating out the ${\tt\bf N}, {\tt\bf \bar N }$ preons
if the tree level superpotential includes the terms
\begin{equation}W_{\rm tree}\supset M_N {\tt\bf \bar N}{\tt\bf N} +\kappa^d
{\tt\bf d} {\tt\bf
\bar N} \bar d  +\lambda^H {\tt\bf h} {\tt\bf N}  {\bar H} \ .
\end{equation}
\subsection{The Light Quark Masses}
In order to give the charm and up quarks mass, it is necessary that the model be
triplicated--that is two more confining $SU(2)$ groups, which get strong at
scales $\Lambda_1$ and $\Lambda_2$  respectively,   produce two more sets of 15
light composite particles. These include the first and second quark 
doublets, the
$\bar u$ and the $\bar c$, with dynamical couplings to two additional 
composite up-type
Higgses. The latter, as well as the additional $D$ and $E$ particles,
will combine with elementary particles to get
large masses of order
$\Lambda_{1,2}$, however off-diagonal superpotential couplings to the down-type
Higgses will cause the heavy up-type Higgses to mix with the lightest 
$H$ by  amounts
of order
$\Lambda_3/\Lambda_2$, $\Lambda_3/\Lambda_1$. The light up-type Higgs is
actually a mixture of the three composite Higgses--which explains the charm and
up masses as dynamical effects. The compositeness scale for the second family
quarks is $>200$~TeV---those of you who are refugees from extended technicolor
model building will recognize this scale as being high enough to keep the
model safe from overly large $K-\bar K$ mixing.

The down and strange quark masses arise in a manner similar to the bottom quark
mass. The number of doublets for each of the $SU(2)$'s is chosen such that if
all the $SU(2)$ couplings are equal at short distance and   the differance in
confinement scales is due entirely to different masses $M_{1,2,3}$ for the three
sets of heavy preons, we obtain the natural order of magnitude relations for
quark masses and mixing angles
\begin{equation}\begin{array}{c} 
 m_d/m_s\sim\sqrt{m_u/m_c}\sim
\theta_{12}\sim\left({M_2/M_1}\right)^{(1/3)}\\
 m_s/m_b\sim\sqrt{m_c/m_t}\sim
\theta_{23}\sim\left({M_3/M_2}\right)^{(1/3)}\\
 \theta_{13}\sim (M_3/M_1)^{1/3}\ .
\end{array}\end{equation}
The $M_{i}$'s can be chosen such that these all work to within a factor of 2 or
3.

\subsection{Lepton Masses}
Since the leptons and the down-type Higgs are both fundamental particles,
renormalizable lepton-Higgs  couplings are allowed. The lepton mass hierarchy
could be put in by hand. However work is in progress on an attempt to explain the
lepton Yukawa coupling hierarchy via large anomalous dimensions, induced by
a superpotential coupling of the lepton doublets to the strongly coupled preons
${\tt\bf h},{\tt\bf N'}, {\tt\bf
\bar N'}$ \cite{progress}. (This can be done in a way consistent with baryon
and lepton number symmetries.)

\subsection{Supersymmetry and Electroweak Symmetry Breaking}
 A low scale for the messenger sector is preferred in this
model. If the supersymmetry breaking is communicated above the compositeness
scale of the second family, (as in hidden sector models with supergravity as the
messenger,) then strong renormalization effects will ensure that the first two
generation squark masses are not  degenerate. Furthermore
the squark masses will align with the up-type rather than the down-type
quark masses. Thus unless the first two families of squarks are rather heavy,
hidden sector supersymmetry breaking will necessarily lead to overly
large $K-\bar K$  mixing.

A supersymmetry breaking sector such as one of the
gauge mediated DSB models of ref.~\cite{dnns,dns} can easily be appended to this
model, resulting in a
  realistic picture with no large flavor changing neutral currents.

As usual the Higgs potential and electroweak symmetry breaking is determined by 
the supersymmetry breaking sector. However, unlike in the
usual gauge mediated scenario, the up-type Higgs is a composite, and its
supersymmetry breaking  mass is not easily predicted. It is
concievable that $\tan\beta$ could be less than $1$, even in a gauge mediated
scenario.

\subsection{Experimental Tests}
With a sufficiently low third family and Higgs compositeness scale,
\cite{composite,progress} detectable deviations from the standard model could
be found in the $\rho $ parameter, $B-\bar B$ mixing and CP violation, the
$Z\rightarrow b
\bar b$ rate, and the rate and  the lepton distributions and polarization for
$b\rightarrow s \ell^+\ell^-$. A remarkable feature of the quindecuplet model
is an approximate $SU(6)$ global symmetry which allows all of these effects to
be predicted in terms of $\tan\beta$ and a single strong interaction
coefficient 
\cite{composite}. 
Future precision measurements of Higgs and top couplings
would also show small deviations from the standard model.
\section{Conclusions}
There still remains  much exploration to do of strongly coupled supersymmetric
gauge theories. Both dynamical supersymmetry breaking and the
fermion mass hierarchy could potentially be explained with new strong
interactions. It is encouraging that construction of realistic supersymmetric
composite models is possible.  Although I have no example, 
it is especially tempting to speculate
that the same new strong interactions could account for both the gauge and 
flavor
hierarchies---that dynamical supersymmetry breaking will occur in some 
(yet to be discovered) composite
model of quarks and leptons which also sheds light on the flavor puzzle.

\end{document}